\documentclass[superscriptaddress,preprint]{revtex4}
\usepackage{mathrsfs}
\usepackage{amssymb}
\usepackage[tbtags]{amsmath}
\usepackage{graphicx}
\usepackage{epsfig,graphicx,times}
\usepackage{color}
\usepackage{CJK}

\begin{document}

\title{Distinguishing left- and right-handed molecules by two-step coherent
pulses}
\author{W. Z. Jia}
\affiliation{Quantum Optoelectronics Laboratory, Southwest Jiaotong
University, Chengdu
610031, China\\
}
\author{L. F. Wei}
\email{weilianfu@gmail.com, lfwei@home.swjtu.edu.cn}
\affiliation{Quantum Optoelectronics Laboratory, Southwest Jiaotong
University, Chengdu
610031, China\\
}
\date{\today }

\begin{abstract}
Chiral molecules with broken parity symmetries can be modeled as
quantum systems with cyclic-transition structures. By using these
novel properties, we design two-step laser pulses to distinguish
left- and right-handed molecules from the enantiomers. After the
applied pulse drivings, one kind chiral molecules are trapped in
coherent population trapping state, while the other ones are pumped
to the highest states for ionizations. Then, different chiral
molecules can be separated.
\end{abstract}

 \maketitle

\bigskip




\bigskip


Chiral molecule lacks an internal plane of symmetry and consequently
is not superposable on its mirror image. The coexistence of left-
and right-handed chiral molecules (called \textquotedblleft
enantiomers\textquotedblright)
originates from the fundamental broken symmetries in nature~\cite{chirality}%
. The physiological effect of enantiomers of biologically active
compounds may differ significantly \cite{differLR}. In general, only
one enantiomeric form has the potential to be biologically
beneficial, while the other one could be harmful or fatal. Thus,
chiral purification and discrimination of enantiomers are
fundamentally important tasks in pharmacology, biochemistry,
etc~\cite{importance}.
\begin{figure}[b]
\includegraphics[width=0.5\textwidth]{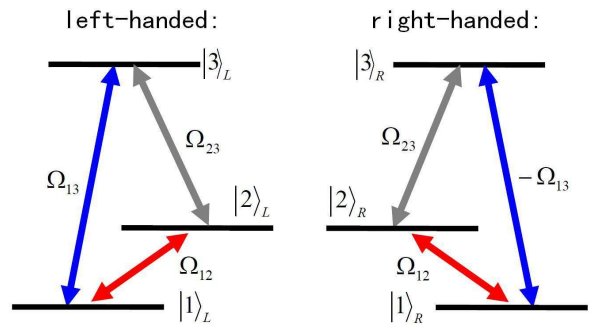}
\caption{(Color online) Three-level chiral molecules with cyclic
transition structures. The enantiomers are resonantly driven by
three optical fields
with Rabi frequencies $\Omega _{12}$, $\Omega _{23}$, and $\pm \Omega _{13}$%
, respectively.} \label{CMEL}
\end{figure}
Traditionally, various chromatographic separation methods, e.g., gas
chromatography (GC), high performance liquid chromatography (HPLC),
gas
liquid chromatography (GLC) and thin layer chromatography (TLC) and so on~%
\cite{Chrom}, are widely utilized to achieve enantioseparation.
Recently, certain optical
means~\cite{Shapiro91,Salam98,Fujimura99,Shapiro2000} were also
proposed to implement the desirable purification. In particular,
these chiral molecules are quantum systems with broken-parity
symmetries and the usual electric-dipole selection rules do not
hold. As a consequence, if only the three lowest levels are
considered, a chiral molecule (left-handed or right-handed) can be
treated as a quantum system with $\Delta$-type cyclic-transition
structure. Note that cyclic transitions can also be realized in some
other quantum systems, e.g., superconducting artificial
atoms~\cite{ArtificialAtom1,ArtificialAtom2}. Recent studies show
that based on this novel property, by coupling these
cyclic-transition molecules with
optical fields can offer some new ways, e.g., cyclic population transfer~%
\cite{Shapiroprl87,Shapiroprl90}, generalized Stern-Gerlach effect~\cite%
{LYprl}, dynamical control~\cite{LYpra}, to achieve
enantioseparation.

In this paper, we develop a dynamical method by using only two-step
ultrashort coherent pulses to implement the desirable
enantioseparation. Due to the fact that only dynamical
ultrashort-pulse operations are used, dynamical method will work
faster and make decoherence effects less important~\cite{LYpra}. In
our protocol, both cyclic transition and coherent population
trapping (CPT)~\cite{CPTArimondo1,CPTGray,CPTArimondo2} are utilized
to achieve enantioseparation.
A specific example of the chiral molecules with cyclic transition,
e.g., is the $D_{2}S_{2}$ enantiomers, as shown in
Ref.~\cite{Shapiroprl90}.
Our basic idea is as follows: First, each enantiomer is prepared at
the superposition of the two relatively lower states by applying an
ultrashort $\pi/2$ pulse. Subsequently, two pulses are
simultaneously applied to couple these lower states to the highest
one. Suppose that the Rabi frequencies of the pulses are chosen
appropriately, so that for one kind chiral molecules, the state
prepared in the first-step pulse operation can be just CPT state
during the second-step pulses, while the other kind chiral molecules
are pumped to the highest quantum state after a $\pi$ rotation.
Then, two kinds of chiral molecules can be separated by using
ionization, followed by ions extraction by an electric field as
suggested in Refs.~\cite{Shapiroprl87,LYpra}.

\begin{figure}[t]
\includegraphics[width=0.5\textwidth]{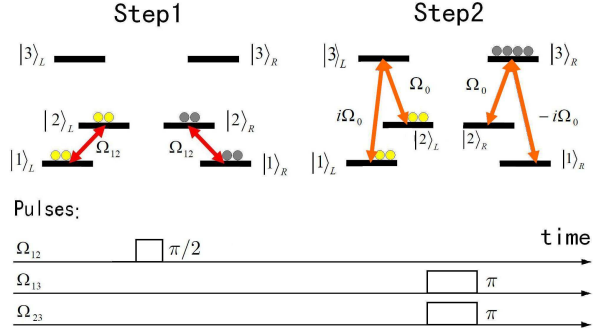}
\caption{(Color online) Schematic representation of the two-step
pulse process to discriminate mixture of molecules. Step 1: Both
left- and right-
handed molecules are prepared in superpositions of their lower states $%
\left\vert \protect\psi _{1}\right\rangle _{L,R}=\left( \left\vert
1\right\rangle _{L,R}-i\left\vert 2\right\rangle _{L,R}\right) /\protect%
\sqrt{2}$ by a $\protect\pi /2$ pulse. Step 2: Two pluses are
simultaneously applied to maintain the state $\left\vert
\protect\psi _{1}\right\rangle _{L} $ invariant, but generate a
$\protect\pi $-rotation transferring the population in the state
$\left\vert \protect\psi _{1}\right\rangle _{R}$ completely to the
state $-\left\vert 3\right\rangle _{R}$ for ionizations}
\label{Pulse}
\end{figure}

%
The transition structures of chiral molecules considered in this
paper are schematized in Fig.~1. Here, only three lowest levels in
the chiral molecules are considered, and thus each chiral molecule
can be modeled as a
three-level cyclic quantum system \cite%
{Shapiroprl87,Shapiroprl90,LYprl,LYpra}. Three laser beams are
applied to drive the enantiomeric molecules. The Hamiltonian of
system can be written as
\begin{equation}
H=\sum_{i=1}^{3}E_{i}\left\vert i\right\rangle \left\langle
i\right\vert +\sum_{j>i=1}^{3}\left[ \Omega_{ij}e^{-i\omega
_{ji}t}\left\vert j\right\rangle \left\langle i\right\vert
+\mathrm{h.c.}\right]
\end{equation}
where $E_{i}$ are the eigenvalues of energy eigenstates $\left\vert
i\right\rangle $. $\omega _{ij}$ are the frequencies, and $\Omega
_{ij}$ the Rabi frequencies of the applied coherent driving fields.
Note that $\Omega _{ij}$ should be regarded as complex parameters
for the optical responses of
quantum systems with loop transition structures are phase sensitive~\cite%
{Shapiroprl87,Shapiroprl90,LYprl,LYpra,loopatom}. Let $\Delta
_{1}=E_{3}-E_{1}-\omega _{31}$, $\Delta _{2}=E_{2}-E_{1}-\omega _{21}$ and $%
\Delta _{3}=E_{3}-E_{2}-\omega _{32}$ be the detunings of the
applied driving fields. When the resonant condition $\Delta _{i}=0$
is satisfied, in interaction picture, the interaction Hamiltonian
can be written as:
\begin{equation}
H_{I}=\sum_{j>i=1}^{3}\left[ \Omega _{ij}\left\vert j\right\rangle
\left\langle i\right\vert +\mathrm{H.c.}\right] .  \label{Hi}
\end{equation}
One of a typical features in the present cyclic transition systems
is that depending on the polarizations of the fields, one or all of
the three Rabi frequencies $\Omega _{ij}$ for the two kinds of
chiral molecules differ by a sign~\cite{Shapiroprl87,Shapiroprl90},
i.e., the total phases of the three Rabi frequencies differ by $\pi
$ between the enantiomers. This means that the left- and
right-handed molecules could be distinguished by a phase-dependent
dynamical process applied to the enantiomers. Following the
Refs.~\cite{Shapiroprl90,LYpra}, the Rabi frequencies for driving
the left- and right-handed molecules are chosen as, $\Omega
_{ij}^{L}\left( t\right) =\Omega _{ij}\left( t\right) $, $\Omega
_{13}^{R}\left( t\right) =-\Omega _{13}\left( t\right) $, $\Omega
_{12}^{R}\left( t\right) =\Omega _{12}\left( t\right) $, $\Omega
_{23}^{R}\left( t\right) =\Omega _{23}\left( t\right) $. We assume
the initial states of left- and right-handed molecules are their
ground states $\left\vert 1\right\rangle _{L}$ and $\left\vert
1\right\rangle _{R}$, respectively. To achieve the desirable
enantioseparation, the left- and right-handed molecules should be
engineered at the quantum states with different energies. To do
this, we design the following two-step pulse process shown in
Fig.~\ref{Pulse}.

Firstly, a $\pi /2$ pulse with Rabi frequency $\Omega _{12}\left(
t\right) =\Omega _{12}^{L}\left( t\right) =\Omega _{12}^{R}\left(
t\right) $ is applied to the mixture of chiral molecules for
producing the superpositions of the lowest two energy-states. In
this step, $\Omega _{12}\left( t\right) $ is set as real parameter,
i.e. its phase is zero. During this pulse the interaction
Hamiltonians for the left- and right-handed molecules read
\begin{equation*}
H_{1}^{L}=\Omega _{12}\left\vert 2\right\rangle _{LL}\left\langle
1\right\vert +\mathrm{H.c.}
\end{equation*}
and
\begin{equation*}
H_{1}^{R}=\Omega _{12}\left\vert 2\right\rangle _{RR}\left\langle
1\right\vert +\mathrm{H.c.},
\end{equation*}
respectively. Clearly, the chiral molecules undergo the dynamical
evolutions
\begin{eqnarray}
\left\vert \Psi _{1}(t)\right\rangle _{L,R} &=&\cos \left(
\int_{0}^{t}\Omega _{12}d\tau \right) \left\vert 1\right\rangle
_{L,R}-i\sin \left( \int_{0}^{t}\Omega _{12}d\tau \right) \left\vert
2\right\rangle _{L,R}.  \label{evolution1}
\end{eqnarray}
Obviously, if the duration of the applied pulse is designed as $%
\int_{0}^{t}\Omega _{12}d\tau =\pi /4$, namely a $\pi /2$ pulse, the
states of the enantiomers are prepared as $\left\vert \psi
_{1}\right\rangle _{L}=\left( \left\vert 1\right\rangle
_{L}-i\left\vert 2\right\rangle _{L}\right) /\sqrt{2}$ and
$\left\vert \psi _{1}\right\rangle _{R}=\left(
\left\vert 1\right\rangle _{R}-i\left\vert 2\right\rangle _{R}\right) /\sqrt{%
2}$ respectively.

Secondly, we apply simultaneously two pump pulses coupling transitions $%
\left\vert 1\right\rangle_{L,R}\leftrightarrow \left\vert
3\right\rangle _{L,R}$ and $\left\vert 2\right\rangle
_{L,R}\leftrightarrow \left\vert 3\right\rangle _{L,R}$ with
properly designed Rabi frequencies for evolving the different chiral
molecules into different quantum states.

The Rabi frequencies of the second-step pulses could be designed as
\begin{eqnarray}
\Omega _{13}^{L}\left( t\right) &=&-\Omega _{13}^{R}\left( t\right)
=\Omega
_{13}\left( t\right) =i\Omega _{0}\left( t\right) ,  \notag \\
\Omega _{23}^{L}\left( t\right) &=&\Omega _{23}^{R}\left( t\right)
=\Omega _{23}\left( t\right) =\Omega _{0}\left( t\right) ,
\label{Rabi2a}
\end{eqnarray}
with $\Omega _{0}\left( t\right) =$ $\left\vert \Omega _{23}\left(
t\right) \right\vert =\left\vert \Omega _{13}\left( t\right)
\right\vert .$The according Hamiltonians can be written as
\begin{eqnarray}
H_{2}^{L} &=&\Omega ^{^{\prime }}\left\vert 3\right\rangle
_{LL}\left\langle
\Phi \right\vert +\mathrm{H.c.}  \label{HL2} \\
H_{2}^{R} &=&\Omega ^{^{\prime }}\left\vert 3\right\rangle
_{RR}\left\langle \Phi \right\vert +\mathrm{H.c.}  \label{HR2}
\end{eqnarray}
with $\left\vert \Phi \right\rangle _{L}=\left( -i\left\vert
1\right\rangle _{L}+\left\vert 2\right\rangle _{L}\right)
/\sqrt{2}$, $\left\vert \Phi \right\rangle _{R}=\left( i\left\vert
1\right\rangle _{R}+\left\vert
2\right\rangle _{R}\right) /\sqrt{2}$ and the effective Rabi frequency $%
\Omega ^{^{\prime }}=\sqrt{2}\Omega _{0}$. It is easily checked that
the state $\left\vert \psi _{1}\right\rangle _{L}$ is a CPT state
for the Hamiltonian \eqref{HL2} with zero eigen value (Note that
$\left\vert \psi _{1}\right\rangle _{L}$ is not belong to the
subspace spanned by $\left\vert \Phi \right\rangle _{L}$ and
$\left\vert 3\right\rangle _{L}$.). This means the state $\left\vert
\psi _{1}\right\rangle _{L}$ is unchanged after this operation,
i.e., $\left\vert \psi _{2}\right\rangle _{L}=\left\vert \psi
_{1}\right\rangle _{L}$. While, for the right-handed molecules
prepared in the state $\left\vert \psi _{1}\right\rangle _{R}$ after
the first step pulse, the dynamical evolution during this operation
reads
\begin{equation}
\left\vert \Psi _{2}(t)\right\rangle _{R}=-i\cos \left(
\int_{0}^{t}\Omega ^{^{\prime }}dt\right) \left\vert \Phi
\right\rangle _{R}-\sin \left( \int_{0}^{t}\Omega ^{^{\prime
}}dt\right) \left\vert 3\right\rangle _{R}. \label{evolution2}
\end{equation}
Obviously, after a $\pi $ rotation for the effective Rabi frequency
$\Omega ^{^{\prime }}$, i.e., $\int_{0}^{t}\Omega ^{^{\prime
}}dt=\pi /2$, the right-handed molecules evolve to $\left\vert \psi
_{2}\right\rangle _{R}=-\left\vert 3\right\rangle _{R}$. Note that
during this process, the left-handed molecules always maintain the
dark state $\left( \left\vert 1\right\rangle _{L}-i\left\vert
2\right\rangle _{L}\right) /\sqrt{2}$, as mentioned above.

\begin{figure}[t]
\includegraphics[width=0.7\textwidth]{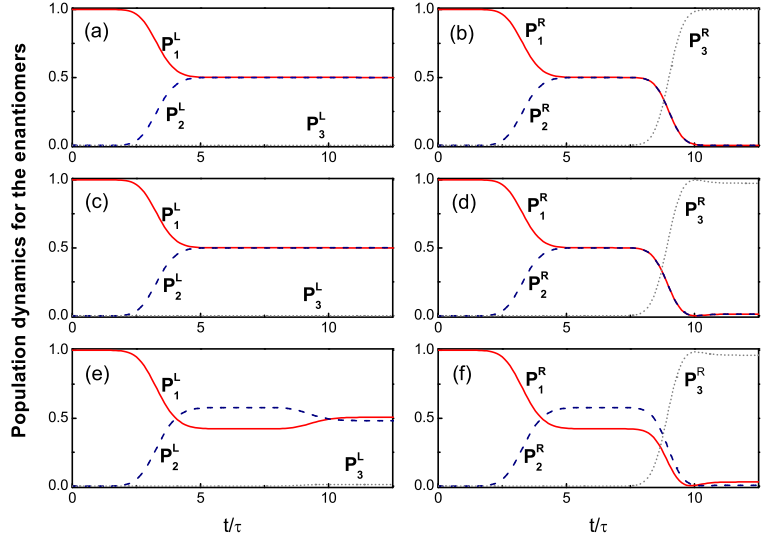}
\caption{(Color online) The time evolution of the population of the
chiral molecules. Both enantiomers start in their ground states
$\left\vert 1\right\rangle _{L,R}$. Subplots (a) and (b): The two
step-pulses are accurate. At the end of the process, the left-handed
molecules are completely trapped in $\left\vert 1\right\rangle _{L}$
and $\left\vert 2\right\rangle _{L}$ states, while the right-handed
molecules are totally pumped to $\left\vert 3\right\rangle _{R}$
state. Subplots (c) and (d): The pulse duration in the second step
are inaccurate. After pulse operation, the
left-handed molecules are not influenced and completely trapped in $%
\left\vert 1\right\rangle _{L}$ and $\left\vert 2\right\rangle _{L}$
states, while the right-handed molecules are partially pumped to
$\left\vert 3\right\rangle _{R}$ state. Subplots (e) and (f): All
the pulses, including their durations and the relative phase\ of the
two pulses used in the second step, are inaccurate (Here we assume
all these quantities are with $10\%$ deviation from idea values.).
After pulse operation, the left-handed molecules are still mostly
populated in $\left\vert 1\right\rangle _{L}$ and $\left\vert
2\right\rangle _{L}$ states, and the right-handed molecules are
nearly pumped to $\left\vert 3\right\rangle _{R}$ state. This shows
that the two-step pulse approach is robust, even the operations are
imperfect.} \label{PD}
\end{figure}

Figs.~\ref{PD} (a) and (b) give the the time evolution of
populations $P_{i}^{L,R}\left( t\right) $ in states $\left\vert
i\right\rangle _{L,R}$ for each kind chiral molecules under accurate
two-step pulses. In our simulation, the applied Rabi frequencies are
designed as following: in the first step, we start with a $\pi /2$\
pulse couples levels $\left\vert 1\right\rangle_{L,R}$ and
$\left\vert 2\right\rangle_{L,R}$, and has Rabi frequency
$\Omega_{12}\left( t\right)=\frac{\sqrt{\pi }}{4\tau }\mathrm{Exp
}[-\left( t-3\tau \right) ^{2}/\tau ^{2}]$;
and in the second step, we simultaneously add two pulses with the
Rabi frequencies
$\Omega _{13}\left( t\right) =\frac{i}{2\tau }\sqrt{\frac{\pi
}{2}}\mathrm{Exp}[-\left( t-9\tau \right) ^{2}/\tau ^{2}]$
and
$\Omega_{23}\left( t\right) =\frac{1}{2\tau} \sqrt{\frac{\pi
}{2}}\mathrm{Exp}[-\left( t-9\tau \right) ^{2}/\tau ^{2}]$
 that
couple $\left\vert 1\right\rangle_{L,R}$ $\leftrightarrow \left\vert
3\right\rangle _{L,R}$ and $\left\vert 2\right\rangle
_{L,R}\leftrightarrow \left\vert 3\right\rangle _{L,R}$,
respectively (Note that $\Omega _{13}\left( t\right) $ and $\Omega
_{23}\left( t\right) $ can generate an effective $\pi $ rotation
between states $\left\vert \Phi \right\rangle _{L,R}$ and
$\left\vert 3\right\rangle _{L,R}$ with Rabi frequency
$\Omega ^{\prime }\left( t\right) =\frac{\sqrt{\pi }}{2\tau
}\mathrm{Exp}[-\left( t-9\tau \right) ^{2}/\tau ^{2}]$
.). If the left- and right-handed molecules are initially prepared
in their ground states $\left\vert 1\right\rangle _{L}$ and
$\left\vert 1\right\rangle _{R}$, after this accurate two-step pulse
process, the final populations of left- and
right-handed molecules are $p_{1}^{L}= p_{2}^{L}=1/2$, $p_{3}^{L}=0$ and $%
p_{1}^{R}= p_{2}^{R}=0$, $p_{3}^{R}=1$, as shown in Figs.~\ref{PD}
(a) and (b).

Given the two kinds of chiral molecules evolving to different states
with different energies after the above two-step pulses, an
ionization process, as suggested in Ref.~\cite{Shapiroprl87}, can be
utilized to separate them. Specifically, in our case, by appropriate
ionization energies, the right-handed molecules prepared in the
highest levels can be ionized but the left-handed ones populating in
the superposition of two lower states remain unchanged. Followed by
the ions extractions via an electric field, the mixture of
enantiomers can be finally separated.

Practically, in the two-step pulse process, the Rabi frequencies,
either the intensities or the phases, may be inaccurate. Now we
begin to discuss the influence of all these deviations on the
enantioseparation. At first, we assume that in the first step, the
applied pulse deviate from perfect $\pi/2$ pulse and according Rabi
frequencies then could be written as $\Omega _{12}^{L}=\Omega
_{12}^{R}=\Omega _{12}=\widetilde{\Omega }_{12}+\delta
\widetilde{\Omega }_{12}$, with $\int_{0}^{t}\widetilde{\Omega
}_{12}dt=\frac{\pi }{4}$, $\int_{0}^{t}\delta \widetilde{\Omega}
_{12}dt=\Delta $. Clearly, after this pulse operation, the states of
the enantiomers are prepared as
\begin{equation}
\left\vert \psi_{1}^{\prime}\right\rangle _{L,R}=\cos \Delta \frac{1}{\sqrt{2}}%
\left( \left\vert 1\right\rangle _{L,R}-i\left\vert 2\right\rangle
_{L,R}\right) -\sin \Delta \frac{1}{\sqrt{2}}\left( \left\vert
1\right\rangle _{L,R}+i\left\vert 2\right\rangle _{L,R}\right) .
\label{state1afterIP}
\end{equation}
In addition, we assume that in the second step, the simultaneously
applied two pump pulses are also imperfect ones with Rabi
frequencies
$\Omega _{13}^{L}\left( t\right) =-\Omega _{13}^{R}\left( t\right)
=\Omega _{13}\left( t\right) =e^{i\left( \frac{\pi }{2}+\delta \phi
\right) }\left( \widetilde{\Omega }_{0}+\delta \widetilde{\Omega
}_{0}\right)$,
$\Omega _{23}^{L}\left( t\right) =\Omega _{23}^{R}\left( t\right)
=\Omega _{23}\left( t\right) =\widetilde{\Omega }_{0}+\delta
\widetilde{\Omega }_{0}$
, where $\delta \widetilde{\Omega}_{0}$ and $\delta \phi$ represent
the deviations of amplitude and phase of Rabi frequency. Let
$\Omega^{\prime }= \widetilde{\Omega }^{\prime }+\delta
\widetilde{\Omega }^{\prime }=\sqrt{2} \left( \widetilde{\Omega
}_{0}+\delta \widetilde{\Omega }_{0}\right) $,
the effective Hamiltonian describing this step can be written as
\begin{eqnarray}
H_{2} ^{\prime L} &=&\Omega ^{\prime }\left\vert 3\right\rangle
_{LL}\left\langle
\Phi ^{\prime }\right\vert +\mathrm{H.c.}  \label{IPHL2} \\
H_{2} ^{\prime R} &=&\Omega ^{\prime }\left\vert 3\right\rangle
_{RR}\left\langle \Phi ^{\prime }\right\vert +\mathrm{H.c.}
\label{IPHR2}
\end{eqnarray}
with
$\left\vert \Phi ^{\prime }\right\rangle
_{L}=\frac{1}{\sqrt{2}}\left( -ie^{-i\delta \phi }\left\vert
1\right\rangle _{L}+\left\vert 2\right\rangle _{L}\right)$,
$\left\vert \Phi ^{\prime }\right\rangle _{R}=\frac{1}{\sqrt{2%
}}\left( ie^{-i\delta \phi }\left\vert 1\right\rangle
_{R}+\left\vert 2\right\rangle _{R}\right)$.
We can define the states
$\left\vert \varphi ^{\prime }\right\rangle
_{L}=\frac{1}{\sqrt{2}}\left( \left\vert 1\right\rangle
_{L}-ie^{i\delta \phi }\left\vert 2\right\rangle _{L}\right)$
and
$\left\vert \varphi ^{\prime }\right\rangle _{R}=\frac{1}{\sqrt{2}}%
\left( \left\vert 1\right\rangle _{R}+ie^{i\delta \phi }\left\vert
2\right\rangle _{R}\right)$,
which are orthogonal to $\left\vert \Phi ^{\prime }\right\rangle
_{L}$ and $\left\vert \Phi ^{\prime }\right\rangle _{R}$,
respectively. Thus, the initial states of left- and right-handed
molecules of the second pulse operation can be rewritten as
\begin{equation}
\left\vert \psi_{1}^{\prime}\right\rangle _{L,R}=A_{L,R}\left\vert
\Phi ^{\prime }\right\rangle _{L,R}+B_{_{L,R}}\left\vert \varphi
^{\prime }\right\rangle _{L,R} \label{Initial state of step2 IP}
\end{equation}
with
$A_{L,R}=\frac{1}{2}i\left[ \pm \cos \Delta \left( e^{i\delta \phi
}\mp 1\right) \mp \sin \Delta \left( e^{i\delta \phi }\pm 1\right)
\right]$,
$B_{L,R}=\frac{1}{2}\left[ \cos \Delta \left( 1\pm e^{-i\delta \phi
}\right) -\sin \Delta \left( 1\mp e^{-i\delta \phi }\right) \right]
$.
Obviously, in the second pules process, the dynamical evolution of
the chiral molecules reads
\begin{equation}
\left\vert \Psi_{2}^{\prime}\left( t\right) \right\rangle
_{L,R}=A_{L,R}\cos \left( \int_{0}^{t}\Omega ^{\prime }dt\right)
\left\vert \Phi ^{\prime }\right\rangle _{L,R}-iA_{L,R}\sin \left(
\int_{0}^{t}\Omega ^{\prime }dt\right) \left\vert 3\right\rangle
_{L,R}+B_{L,R}\left\vert \varphi ^{\prime }\right\rangle _{L,R}.
\label{evolution step2 IP}
\end{equation}
Clearly, after an imperfect $\pi $ rotation with the effective Rabi
frequency $\Omega ^{\prime }=\widetilde{\Omega }^{^{\prime }}+\delta
\widetilde{\Omega }^{^{\prime }}$ (Here we assume that $\int_{0}^{t}\widetilde{%
\Omega }^{^{\prime }}dt=\frac{\pi }{2}$, $\int_{0}^{t}\delta \widetilde{\Omega }%
^{^{\prime }}dt=\Delta ^{^{\prime }}$.), the left- and right-handed
molecules evolve to the states
\begin{equation}
\left\vert \psi_{2}^{\prime}\right\rangle _{L,R}=-A_{L,R}\sin \Delta
^{\prime }\left\vert \Phi ^{\prime }\right\rangle
_{L,R}-iA_{L,R}\cos \Delta ^{\prime }\left\vert 3\right\rangle
_{L,R}+B_{L,R}\left\vert \varphi ^{\prime }\right\rangle _{L,R}.
\label{final state of step 2 IP}
\end{equation}
As a consequence, the final populations of the enantiomers for the small $%
\Delta $, $\Delta ^{^{\prime }}$, $\delta \phi $ read
\begin{eqnarray}
p_{1,2}^{L} &\thickapprox &\frac{1}{2}\left[ 1-\Delta ^{2}\pm
2\Delta \Delta ^{\prime }-\frac{1}{4}\left( \delta \phi \right)
^{2}\right] ,
\label{final population L12} \\
p_{3}^{L} &\thickapprox &\frac{1}{4}\left( \delta \phi \right)
^{2}+\Delta
^{2},  \label{final population L3} \\
p_{1,2}^{R} &\thickapprox &\frac{1}{2}\left[ \left( \Delta ^{\prime
}\pm \Delta \right) ^{2}+\frac{1}{4}\left( \delta \phi \right)
^{2}\right] ,
\label{final population R12} \\
p_{3}^{R} &\thickapprox &1-\Delta ^{2}-\Delta ^{\prime
2}-\frac{1}{4}\left( \delta \phi \right) ^{2}.  \label{final
population R3}
\end{eqnarray}

Now we give some discussion on the above results. If the states
prepared in the first step are perfect, then, in the second step,
even if inaccurate pulse amplitude (or duration) lead to deviation
of perfect $\pi $ pulses (Here, we assume that Eq.~\eqref{Rabi2a},
i.e., CPT condition $\Omega _{13}\left( t\right) /\Omega _{23}\left(
t\right) =i$ for left-handed molecules, should still be satisfied.),
purification of the right-handed
molecules can also be effectively performed. To show this, we let $%
\Delta =0$, $\delta \phi =0$ in Eq.~\eqref{final population
L12}-\eqref{final population R3} and thus the final populations in
this case are $p_{1,2}^{L}\thickapprox \frac{1}{2}$,
$p_{3}^{L}\thickapprox 0$, $p_{1,2}^{R}\thickapprox
\frac{1}{2}\Delta ^{\prime 2}$, $p_{3}^{R}\thickapprox 1-\Delta
^{\prime 2}$, respectively. This means that the left-handed
molecules can be trapped in a superposition state of the two lower
states, i.e., the CPT state $\left\vert \psi_{1}\right\rangle _{L}$,
while right-handed molecules can still be \textit{partially} pumped
to the higher state under imperfect $\pi$ pulses, and then purified
by the following ionization process. Note that in this case not both
enantiomers can be perfectly purified, as not only left-handed
molecules but also partial right-handed ones populate in the two
lower states ($p_{1,2}^{R}\thickapprox \frac{1}{2}\Delta ^{\prime
2}$). Thus, after ions formed by right-handed molecules extracted by
an electric field, the remainder is still mixture of two
enantiomers. But this is still meaningful, since in most cases only
one enantiomer may produce the desired therapeutic (or biological)
activities and thus we should extract merely the biologically
beneficial chiral molecules from mixtures of enantiomers. To confirm
the above analysis, Figs.~\ref{PD} (c) and (d) present a numerical
simulation of the population dynamics for the left-handed and
right-handed
molecules initially prepared in states $\left\vert 1\right\rangle _{L}$ and $%
\left\vert 1\right\rangle _{R}$ and driven by pulses with Rabi
frequencies
$\Omega _{12}\left( t\right) =\frac{\sqrt{\pi }}{4\tau
}\mathrm{Exp}[-\left( t-3\tau \right) ^{2}/\tau ^{2}]$,
$\Omega _{13}\left( t\right) =\frac{i}{2\tau }\sqrt{\frac{\pi
}{2}}\mathrm{Exp}[-\left( t-9\tau \right) ^{2}/\left( 1.1\tau
\right) ^{2}]$,
and
$\Omega _{23}\left( t\right) =\frac{1}{2\tau}\sqrt{\frac{\pi
}{2}}\mathrm{Exp}[-\left( t-9\tau \right) ^{2}/\left( 1.1\tau
\right) ^{2}]$,
respectively (Note that $\Omega _{13}\left( t\right)$ and $\Omega
_{23}\left( t\right)$ can generate an effective $\pi$ rotation
between states $\left\vert \Phi \right\rangle _{L,R}$ and
$\left\vert 3\right\rangle _{L,R}$ with Rabi frequency
$\Omega ^{\prime }\left( t\right) =\frac{\sqrt{\pi }}{2\tau
}\mathrm{Exp}[-\left( t-9\tau \right) ^{2}/\left( 1.1\right) \tau
^{2}]$,
leading to a deviation of $\pi $ pulse: $\Delta ^{\prime
}\thickapprox 0.1\times \frac{\pi }{2}$.). It can be seen from
Figs.~\ref{PD} (c) and (d) that the left-handed molecules are
finally populated in the two lower states $\left\vert 1\right\rangle
_{L}$ and $\left\vert 2\right\rangle _{L}$ with
$p_{1}^{L}=p_{2}^{L}=1/2$, the right-handed molecules are mostly
populated in state $\left\vert 3\right\rangle _{R}$ with
$p_{3}^{R}\thickapprox 0.976$ and slightly populated in state
$\left\vert 1\right\rangle _{R}$ and $\left\vert 2\right\rangle
_{R}$ with $p_{1}^{R}\thickapprox p_{2}^{R}\thickapprox 0.012$.

For the more general case that all the applied Rabi frequencies in
the two steps, either the intensity or the phase, are inaccurate,
Eq.~\eqref{final population L12}-\eqref{final population R3} give
the final population of enantiomers theoretically. For a numerical
simulation of population dynamics, we assume that the Rabi
frequencies are designed as
$\Omega _{12}\left( t\right) = \frac{\sqrt{\pi }}{4\tau
}\mathrm{Exp}[-\left( t-3\tau \right) ^{2}/\left( 1.1\tau \right)
^{2}]$,
$\Omega _{13}\left( t\right) =\frac{\sqrt{2\pi }}{4\tau
}\mathrm{Exp}[-\left( t-9\tau \right) ^{2}/\left( 1.1\tau \right)
^{2}+i\frac{11\pi }{20}]$
and
$\Omega _{23}\left( t\right) =\frac{\sqrt{2\pi }}{4\tau
}\mathrm{Exp}[-\left( t-9\tau \right) ^{2}/\left( 1.1\tau \right)
^{2}]$,
respectively. Again, note that $\Omega _{12}\left( t\right) $ can
lead to a deviation of $\pi /2$ pulse $\Delta \thickapprox 0.1\times
\frac{\pi }{4};$ $\Omega _{13}\left( t\right) $\ and $\Omega
_{23}\left( t\right) $ can generate an effective rotation between
states $\left\vert \Phi ^{\prime }\right\rangle _{L,R}$ and
$\left\vert 3\right\rangle _{L,R}$ with Rabi
frequency $\Omega ^{\prime }\left( t\right) =\frac{\sqrt{\pi }}{2\tau }%
\mathrm{Exp}[-\left( t-9\tau \right) ^{2}/\left( 1.1\right) \tau
^{2}]$,
leading to a phase imperfection $\delta \phi \thickapprox 0.1\times \frac{%
\pi }{2}$ and a deviation of effective$\ \pi $ pulse $\Delta
^{\prime }\thickapprox 0.1\times \frac{\pi }{2}$. Under these
pulses, the time evolution of population of enantiomers are shown in
Figs.~\ref{PD} (e) and (f). Our numerical calculation shows that
after the pulses given above, the final populations
$p_{1}^{L}+p_{2}^{L}\thickapprox 0.988$, $p_{3}^{R}\thickapprox
0.964$, which is in accordance with theoretical evaluations given by
Eq.~\eqref{final population L12}-\eqref{final population R3}.
Obviously, these results show that if the applied pulses do not
deviate very large from perfect ones, after pulse operation, the
left-handed molecules can be almost populated in states $\left\vert
1\right\rangle_{L}$ and $\left\vert 2\right\rangle_{L}$, and the
right-handed ones can be nearly populated in state $\left\vert
3\right\rangle_{R}$. Moreover, note that in practice the applied
pulses can be more accurate than those used in our simulation and
the influence of this inaccuracy is more less. Thus, even the given
pulses are not so much perfect, the enantiomers can also be
effectively prepared in different final states.

Certainly, during the second driving process, the Rabi frequencies
can also be designed as
\begin{eqnarray}
\Omega _{13}^{L}\left( t\right) &=&-\Omega _{13}^{R}\left( t\right)
=\Omega
_{13}\left( t\right) =-i\Omega _{0}\left( t\right),  \notag \\
\Omega _{23}^{L}\left( t\right) &=&\Omega _{23}^{R}\left( t\right)
=\Omega _{23}\left( t\right) =\Omega _{0}\left( t\right).
\label{Rabi2b}
\end{eqnarray}
In this case, the right-handed molecules are trapped in the dark state $%
|\psi _{1}\rangle_{R}$ and thus unchanged after the pulses, while
the left-handed ones are pumped to the higher state. Thus, after
pulse operation, similar ionization and ions extraction processes
can be implemented to separate the enantiomers.

In conclusion, we have introduced a two-step optical pulses method
to enantioseparations. In our protocol, the pulse processes are
simplified by trapping one kind of chiral molecules in CPT states
and the others are evolved to the highest levels for ionization.
Compared with the previous three-step pulse method~\cite{LYpra}, our
two-step operational approach is more robust in the presence of
decoherence for less pulse steps means shorter operation time.
Additionally, even if the applied pulses are inaccurate, the
enantiomers can also be effectively prepared in different final
states and be separated. Finally, in our method lower ionization
energy is required, as the left(right)-handed molecules are prepared
in their highest states.

\begin{acknowledgments}
The project was supported in part by National Natural Science
Foundation of China under Grant Nos. 10874142, 90921010 and the
National Fundamental Research Program of China through Grant No.
2010CB923104.
\end{acknowledgments}


\end{document}